\documentclass{article}

     \usepackage[final]{neurips_2020_ml4ps}

\usepackage[utf8]{inputenc} 
\usepackage[T1]{fontenc}    
\usepackage{hyperref}       
\usepackage{url}            
\usepackage{booktabs}       
\usepackage{amsfonts}       
\usepackage{nicefrac}       
\usepackage{microtype}      
\usepackage{graphicx}
\usepackage{ulem}

\usepackage{enumitem}
\setlist[itemize]{leftmargin=*}
\usepackage{amssymb, amsmath}
\usepackage{xcolor}

\newcommand{\eg}{e.g.,\ }

\def\eg{{\it e.g.,\ }}

\newcommand{\be} {\begin{equation}}
\newcommand{\ee} {\end{equation}}

\title{Debunking Generalization Error \\ or: How I Learned to Stop Worrying \\ and Love My Training Set} 

\author{%
  Viviana Acquaviva \\
  CUNY NYC College of Technology\\
  300 Jay Street\\
  Brooklyn, NY 11201 \\
  \texttt{vacquaviva@citytech.cuny.edu}
  \And
  Christopher C. Lovell \\
  School of Physics, Engineering and Computer Science\\
  University of Hertfordshire\\
  Hatfield AL10 9AB, UK\\
  \texttt{c.lovell@herts.ac.uk} \\
  \And
  Emille E. O. Ishida \\
 Universit\'e Clermont Auvergne\\
 CNRS/IN2P3, LPC\\ 
 F-63000 Clermont-Ferrand, France \\
\texttt{emille.ishida@clermont.in2p3.fr}\\
  } 
  
\begin{document}

\maketitle

\begin{abstract}

We aim to determine some physical properties of distant galaxies (for example, stellar mass, star formation history, or chemical enrichment history) from their observed spectra, using supervised machine learning methods. We know that different astrophysical processes leave their imprint in various regions of the spectra with characteristic signatures. Unfortunately, identifying a training set for this problem is very hard, because labels are not readily available - we have no way of knowing the true history of how galaxies have formed. One possible approach to this problem is to train machine learning models on state-of-the-art cosmological simulations. However, when algorithms are trained on the simulations, it is unclear how well they will perform once applied to real data.
In this paper, we attempt to model the generalization error as a function of an appropriate measure of distance between the source domain and the application domain. Our goal is to obtain a reliable estimate of how a model trained on simulations might behave on data. 

\end{abstract}

\section{Introduction}

One recurring problem in data-intensive sciences, including Astrophysics, is the fact that collecting and validating large training sets required by supervised machine learning algorithms is very difficult, very expensive, or both. 
To circumvent the difficulty of observing astrophysical processes directly, in the last decade there has been an enormous community effort running large, expensive numerical simulations of galaxy formation and evolution. 
The results of the most sophisticated of these (such as the recent Illustris TNG\footnote{\href{http://www.tng-project.org/}{http://www.tng-project.org/}}) resemble remarkably well the Universe that we presently observe, giving us some confidence that such state-of-the-art simulations are able to capture the most important processes in galaxy formation and evolution.

Recently, it was shown that deep learning algorithms such as Convolutional Neural Networks can be trained to learn galaxy physical properties, including star formation histories, using cosmological simulations, with very promising results \cite{Lovell2019}. It was also shown that the results were relatively robust to training models using one cosmological simulations (Illustris, \cite{genel_introducing_2014}), and then applying it to galaxies from a different simulation (EAGLE, \cite{schaye_eagle_2014}), with the generalization error becoming worse by $\sim$ 25-35\%. 
However, a quantitative assessment of the performance of these models when applied to observed spectra from real galaxies, when there is no ``ground truth" to be known, is still lacking. If one could show that the simulations are realistic enough - in other words, the data sets of simulated and real galaxies are statistically similar - this would go a long way in confirming the reliability of the method. 

\begin{figure}
\centering
	\raisebox{-0.43\height}{\includegraphics[width=0.7\columnwidth]{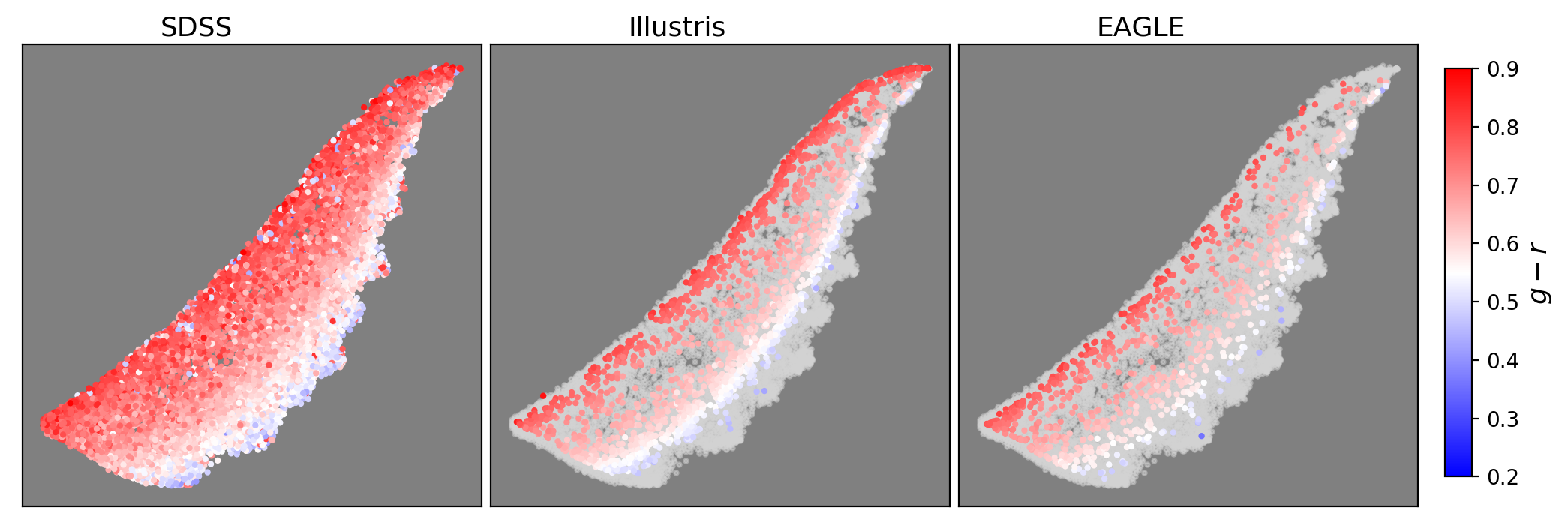}}
    \caption{$t$-SNE plot applied to spectra from a comparable selection of real galaxies in the Sloan Digital Sky Survey (left panel) and the Illustris (middle panel) and EAGLE (right panel) selections. Each point represents a single galaxy spectrum. Nearby points in this 2D space have high spectral similarity. Each distribution is coloured by $g-r$ colour. The SDSS point distribution is shown in the background in light grey for the middle and right panels for comparison. From \cite{Lovell2019}.}
    \label{fig:SMAPE_TSNE}
\end{figure}

One possible test consists of looking at the distribution of three data sets - spectra of real galaxies from SDSS, galaxies from the EAGLE simulations and galaxies from the Illustris simulation, using a dimensionality reduction/projection technique, for example $t$-distributed Stochastic Neighbor Embedding (t-SNE, \cite{maaten2008visualizing}). The results, shown in the right panel of Fig. \ref{fig:SMAPE_TSNE}, are encouraging: the simulations and the data occupy similar spaces, but there are obvious differences (missing points, different densities, etc). Additionally, nonlinear manifold embedding techniques such as t-SNEs require hyperparameter tuning, which makes them subject to overfitting and instability. Our goal is to quantify the effect of such differences more rigorously.

\vspace{-2mm}
\section{Framework}
\vspace{-2mm}
We would like to prove the following hypothesis. We know that in nature, a set of physical properties for galaxies (for example, stellar mass, star formation history, chemical enrichment history...), plus many latent variables, will lead to the spectra that we observe; we could call this ``mapping" function $f(x)$. \textit{We are interested in learning the ``inverse" function $f^{-1}$(x)}, which would teach us to go from observed quantities (spectra) to the quantities that we'd like to measure (physical properties).

In simulations, a set of chosen input physical properties is transformed into a set of simulated spectra by some known modeling function, let's say $g(x)$. Hopefully $g(x)$ is a decent approximation of $f(x)$; we do have some way of verifying this because if our variables are meaningful and our modeling is correct, then the simulated spectra generated by applying $g(x)$ will resemble the observed spectra. In this sense, the distance in spectral space traces the similarity between $f(x)$ and $g(x)$.

Now let us consider the other direction. The function $g^{-1}(x)$ can be learned by, for example, training a machine learning model. The learned function will, of course, have its own generalization error, which depends as usual on a noise term, independent of the model and dependent on degeneracies and data quality, a bias term, and a variance term. This will cause some difference between the ``true" input physical parameters and the inferred physical parameters.

\begin{figure}
\centering
    {\includegraphics[width=0.9\textwidth]{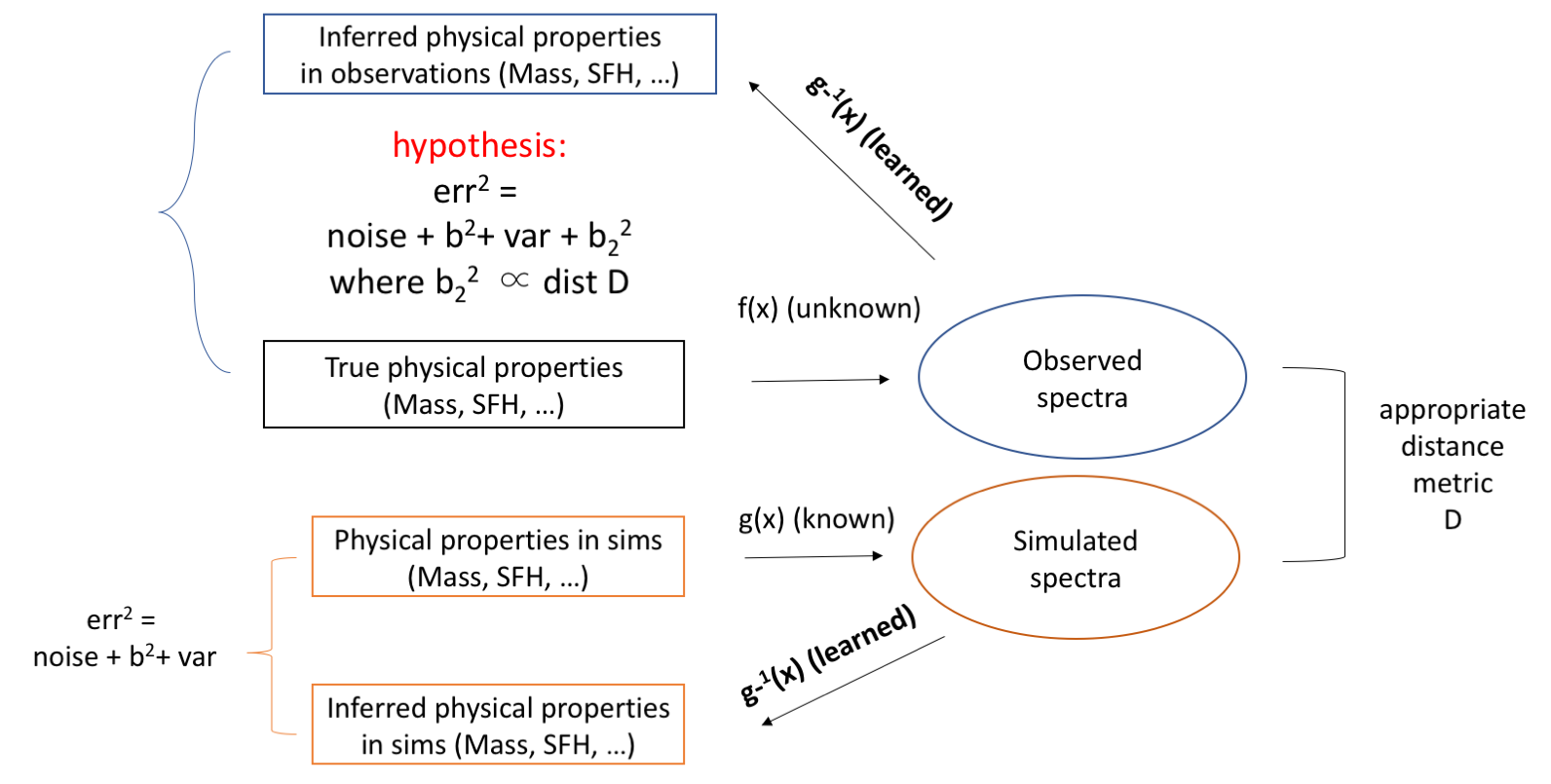}}
    \caption{Scheme of the main hypothesis we are testing: if we can learn an imperfect function $g^{-1}(x)$ that gives us physical properties of galaxies (such as stellar mass, or star formation history, SFH) starting from spectra, can we estimate how much the parameters inferred through  $g^{-1}(x)$ deviate from the true ones, based on some distance D measured in the space of spectra, which acts as a proxy for the distance between $f(x)$ and $g(x)$? We assume that the distance D will act as an additional bias term, $b_2$, in the usual bias/variance/noise decomposition.}
    \label{fig:Scheme1}
\end{figure}

What happens if we apply the learned function $g^{-1}(x)$ \textit{to the observed spectra} (in other words, when we use it a proxy for the function we want, $f^{-1}(x)$)? 
Our hypotheses are the following: 

\vspace{-2mm}
\begin{enumerate}[leftmargin=*]
\itemsep0em 
    \item There is an additional term in the square error, which comes from the fact that we learned the ``wrong" function, $g^{-1}(x)$ instead of $f^{-1}(x)$, which behaves like a bias;
    \item The additional error will depend in a \textit{predictable} way on an appropriate distance metric describing the similarity (or lack of it) between the observed spectra and the simulated spectra. 
\end{enumerate}
\vspace{-2mm}
If we can show that 2. is true, we would be able to predict the generalization error on data. Our scheme is described in Fig. \ref{fig:Scheme1}. 
\vspace{-2mm}
\section{Methodology}\vspace{-2mm}
Our strategy for (empirically) proving our hypotheses consist of these steps:
\begin{enumerate}[leftmargin=*]
\itemsep0em
    \item Generate several sets of simulations, changing the modeling assumptions (in this case, we work with 20 of them, $i = 1-20$). The simulations are chosen to represent the full range of models that are compatible with current observations.
    \item Find a suitable representation feature space for all the simulations sets, and optimize machine learning models in this space;
    \item Identify a simple and appropriate measure of distance between data sets ($D_{i, j}$ where i, j $\in$ [1, 20]);
    \item Train 20 models, one per simulated set of spectra, excluding the objects who participated to the feature selection process in step 2, to learn as many inverse modeling functions, indicated as $g_{1}^{-1}(x), g_2^{-1}(x),... g_{20}^{-1}(x)$;
    \item Apply each of the learned functions to each of the simulated sets of spectra; we note that the spectra to which we apply the learned functions have never interacted with the training process, so these are true test scores;
    \item Generate and analyze 20 scatter plots, one for each simulation set $i$, plotting the distance metric $D_{i, j}$ (where $j = 1, ...20$) versus the generalization error obtained by applying the functions $g_{1}^{-1}(x), g_2^{-1}(x),... g_{20}^{-1}(x)$ to learn the parameters of simulation $i$;
    \item Use these 20 examples to infer a robust regression between the distance metric $D_{i, j}$ mentioned above and the generalization error incurred.
    \item Use the regression model to predict the generalization error on data, based on the distance between data and simulations.
\end{enumerate}

\vspace{-2mm}
\section{Preliminary results}
\vspace{-1mm}
We have generated 20 simulations, each of which contains $\sim$ 6,500 galaxies, with a representative range of physical properties, at redshift $z = 0$. The differences between the simulations arise from using varying modeling assumptions, for example by changing stellar libraries and dust properties. 
The features of this problem are the vector of intensity (brightness) at each wavelength, in the range between 3,000 and 9,000 \AA; this matches the Sloan Digital Sky Survey data. 

We have trained several algorithms, including Random Forests, k Nearest Neighbors, Support Vector Machines, and Convolutional Neural Networks. For this exploratory paper, we have chosen to predict a simple quantity, the Stellar Mass, because of the high interpretability of the problem. At zero order, the stellar mass of a galaxy is proportional to its luminosity, and in particular, to the luminosity in the near-infrared region of the EM spectrum, where fewer confounding effects/degeneracies with other parameters exist. Therefore, we expect to use this problem as a feasibility pilot for more complicated parameter estimation problems.

Our measure of the generalization error is the Mean Square Error (MSE) in stellar mass. Of particular difficulty has been the search for an appropriate distance metric. We have explored many measures of distance between data sets traditionally used in domain adaptation problems, including the Earth Mover Distance of some low-dimensional representation of data such as those resulting from t-SNE maps or Self Organized Maps \cite{SOMs}, the 
Kullback–Leibler divergence of the same maps, the difference in covariance matrices suggested by methods like CORAL \cite{CORAL}, and the Euclidean distance calculated in the space of the Principal Components. However, none of those showed significant correlation with the generalization error. In fact, this problem does not fit the mold of traditional domain adaptation problems, because the domain is the same, but the relationship between input (spectra) and output (physical properties) is different. 

\begin{figure}
\centering
	\includegraphics[width=\textwidth]{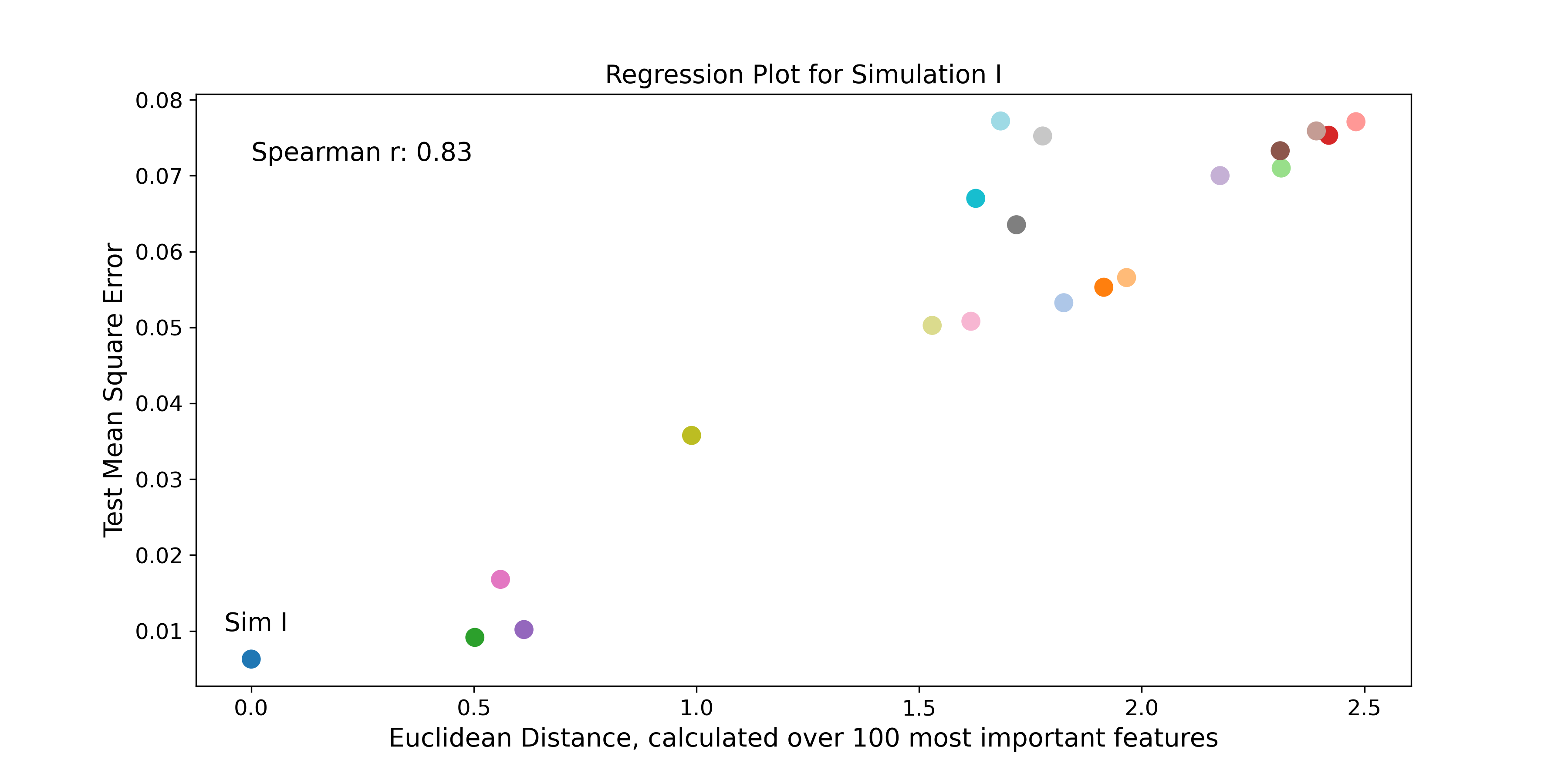}
    \caption{An example scatter plot of Mean Square Error versus pairwise distance, for Simulation 1. The MSE is calculated by applying the inverse mapping function learned on each simulation, from 1 to 20, to the data of simulation 1; each color corresponds to a different simulation. We exclude objects that have participated in the feature selection process. There is a clear correlation trend, which suggest the possibility of successfully fitting a regression model. This would allow us to predict the MSE when the model is applied to data. Plots for the other 19 simulations show similar trends.}
    \label{fig:results}
    \vspace{-0.5cm}
\end{figure}

We have come to the conclusion that a similarity measure that correlates with the MSE in stellar mass, for example, can only be built by using knowledge about which features carry information about the stellar mass. In other words, our metric needs to be problem-specific (unique to each quantity we are going to predict), and derive from a supervised feature selection process, as opposed to an unsupervised dimensionality reduction process.

Our preliminary results are shown for a simplistic feature selection process. We assembled a super-data set by compiling together a random selection of 1000 objects from each of the 20 simulation sets, and fitting a Random Forest Regressor to predict Stellar Mass. We then ranked the features according to their importance, selected the first 100, and calculated the pairwise distance between simulated data sets as the mean Euclidean distance in this 100-dimensional space.

The results are quite promising. We show one example plot where the ``target" set of spectra is simulation 1, and we show the MSE (again from a Random Forest model) when we apply the 20 functions $g_{1}^{-1}, g_2^{-1},... g_{20}^{-1}$ to recover the stellar mass. There is a clear trend that suggest the possibility of fitting the regression successfully. The trends seen here are similar to what we observe in the other 19 plots, where we apply the learned inverse-modeling functions to the other 19 sets of simulations. 
\vspace{-2mm}
\section{Next steps}
\vspace{-2mm}
A great improvement can be achieved quickly by applying a more careful feature selection technique; impurity-based feature selection processes can be quite incorrect when the features are highly collinear, which is the case here. Therefore, we expect that by refining our feature selection technique, for example by clustering highly correlated features and selecting one per cluster, we can obtain better results. 

We also expect that further understanding of the applicability domain of our technique will come from understanding ``failing" cases, such as outliers in our distance/generalization error regressions. 

Additionally, we are aware that the generalization error obtained from tree-based ensemble algorithms, such as that shown in Fig. \ref{fig:results}, can deviate from the expected behavior because of the lack of extrapolation capability of these methods. We expect that using Convolutional Neural Networks, which are already part of our existing framework, will lead to improved results and stability. 

\vspace{-2mm}
\section{Societal Impact}
\vspace{-2mm}
Our findings are very general and could be applied to analog problems in different domains; the issue of applicability of simulations is relevant in many physical disciplines (\eg \cite{Cranmer2019}). Our simulations and models will be available on github, and we hope that this will contribute to advance the discourse on this topic in a transparent manner. On the flip side, one possible concern about negative broader impacts of machine learning research is the carbon footprint of training expensive machine learning models. This is not a cause for concern for the research case presented in this paper, because the size of data sets do not require the use of large scale computational resources, but it is certainly possible that other applications inspired by the work presented here might induce a significant carbon footprint. This should be taken into consideration when discussing costs and benefits of training models repeatedly on large collections of simulations.

\vspace{-2mm}
\bibliographystyle{plainnat}
\bibliography{extracted}

\end{document}